\documentclass[%
 reprint,
superscriptaddress,
 amsmath,amssymb,
 aps,
]{revtex4-2}

\usepackage{graphicx}
\usepackage{dcolumn}
\usepackage{bm}
\usepackage{xcolor}
\usepackage{hyperref}

\renewcommand{\vec}[1]{\mathbf{#1}}

\DeclareMathOperator{\diag}{diag}
\DeclareMathOperator{\Var}{Var}
\DeclareMathOperator{\Cov}{Cov}

\DeclareMathOperator{\logN}{logN}

\begin{document}

\preprint{APS/123-QED}

\title{Non-equilibrium microbial dynamics unveil a new macroecological pattern beyond Taylor's law}

\author{Jos\'e Camacho-Mateu}
\affiliation{%
Grupo Interdisciplinar de Sistemas Complejos (GISC)}
\affiliation{Departamento de Matem\'aticas, Universidad Carlos III de Madrid, Legan\'{e}s, Spain}

\author{Aniello Lampo}
\affiliation{%
Grupo Interdisciplinar de Sistemas Complejos (GISC)}
\affiliation{Departamento de Matem\'aticas, Universidad Carlos III de Madrid, Legan\'{e}s, Spain}

\author{Sa\'ul Ares}
\affiliation{%
Grupo Interdisciplinar de Sistemas Complejos (GISC)}
\affiliation{Centro Nacional de Biotecnologia (CNB), CSIC, Madrid, Spain}%

\author{Jos\'e A. Cuesta}
\email{cuesta@math.uc3m.es}
\affiliation{%
Grupo Interdisciplinar de Sistemas Complejos (GISC)}
\affiliation{Departamento de Matem\'aticas, Universidad Carlos III de Madrid, Legan\'{e}s, Spain}
\affiliation{Instituto de Biocomputaci\'on y F\'{\i}sica de Sistemas Complejos, Universidad de Zaragoza, Zaragoza, Spain}

\date{\today}

\begin{abstract}
We introduce a comprehensive analytical benchmark, relying on Fokker-Planck formalism, to study microbial dynamics in presence of both biotic and abiotic forces. In equilibrium, we observe a balance between the two kinds of forces, leading to no correlations between species abundances. This implies that real microbiomes, where correlations have been observed, operate out of equilibrium. Therefore, we analyze non-equilibrium dynamics, presenting an ansatz for an approximate solution that embodies the complex interplay of forces in the system. This solution is consistent with Taylor’s law as a coarse-grained approximation of the relation between species abundance and variance, but implies subtler effects, predicting unobserved structure beyond Taylor's law. Motivated by this theoretical prediction, we refine the analysis of existing metagenomic data, unveiling a novel universal macroecological pattern. Finally, we speculate on the physical origin of Taylor’s law: building upon an analogy with Brownian motion theory, we propose that Taylor's law emerges as a Fluctuation-Growth relation resulting from equipartition of environmental resources among microbial species.
\end{abstract}

\maketitle

\section{Introduction}

With a global population over $10^{30}$ individuals, microbes exhibit a wider range of physiological diversity than animals and plants combined. They are found across the biosphere, from seawater to the human gut, playing critical roles in ecosystem management and human health. Recent advancements in metagenomics techniques, overcoming traditional laboratory limitations, have revolutionized studies of the microbiome, generating vast datasets. Macroecology has emerged as a valuable approach, employing large-scale patterns to explain microbial abundances and diversity in terms of underlying ecological forces \cite{Jansson2020, Fan2021, Prosser2007, Shoemaker2017, grilli2020macroecological, Ji2020, Zaoli2021}.

The recognition and modeling of these ecological forces are now garnering significant attention. Factors such as environmental effects, cross-feeding, demographic stochasticity, migration, species interactions, and other ecological forces play a pivotal role in shaping microbial communities in both time and space. The existence of such forces is widely acknowledged. However, scholarly research has mainly focused on untangling the effects of these mechanisms to ascertain their relative importance, with limited exploration of their interplay.

Two main perspectives have prevailed. On one hand, there is a viewpoint emphasizing the role of \textit{abiotic} factors, typically captured in a so-called diffusion matrix. This matrix parameterizes fluctuation strengths, modeling the influence of resources and environmental conditions. In particular, it has been shown that environmental fluctuations are responsible for the variation in species abundances across metagenomic samples \cite{grilli2020macroecological} and facilitate the recovery of specific phylogenetic properties \cite{sireci2023environmental}. On the other hand, considerable attention has been directed towards \textit{biotic} interactions, typically represented by an interaction matrix, which underpins the emergence of patterns involving multiple species, such as correlations between fluctuations in the abundances of different species \cite{Ho2022,Camacho2024}.

Despite numerous studies examining biotic and abiotic mechanisms separately, the complex behavior of real ecosystems calls for an integrated approach that considers both of them simultaneously.
We do so through a stochastic Lotka-Volterra model \cite{Camacho2024},
presenting an extended analytical treatment of the system dynamics in terms of the Fokker-Planck equation (FPE), which describes the probability distribution of abundances. We distinguish between equilibrium and non-equilibrium regimes. In equilibrium, the magnitudes of biotic and abiotic forces are balanced, implying vanishing abundance correlations. As nontrivial correlations are observed in real microbiomes, we conclude that microbial dynamics operate out of equilibrium.

The analytical study of non-equilibrium dynamics is challenging though. The intricate interplay between abiotic and biotic forces induces complex configurations of associated flows.
We partly solve this problem by proposing an ansatz that yields an approximate solution when the system is out of equilibrium. To first order, we show that a mean-field formalism recovers the characteristics of this solution, providing arguments for a physical interpretation of the approximation \cite{MN2,Bunin2017,garcia_lorenzana2024}.

Our analysis sheds light on the effective formulation of Taylor's law \cite{Taylor1961}, a ubiquitous ecological pattern spanning a broad spectrum of ecosystems, from microorganisms to complex communities of animals and plants. Taylor's law posits that the variance in species abundance scales as a power law of its mean. In particular, the exponent of such a power-law relationship is two for microbial ecosystems  \cite{grilli2020macroecological}. However, our findings go beyond this coarse-grained description, predicting that the variance-mean relationship across species follows a non-trivial distribution. We confirm this prediction refining the analysis of existing metagenomic data, describing a new universal macroecological pattern beyond Taylor's law. The new distribution observed is valid across several orders of magnitude, and our approach captures the features of this distribution, providing a more nuanced perspective on ecological dynamics.

Finally, to close our discussion on Taylor's law, we make use of an analogy with the physics of Browinian motion to observe that Taylor's law implies for the microbial community a Fluctuation-Growth relation, akin to the Fluctuation-Dissipation found for a Brownian particle when energy equipartition is considered. This leads us to speculate on the physical origin of Taylor's law as some sort of equipartition of environmental resources between different microbial species.

The manuscript is organized as follows. In Section \ref{sec:model}, we introduce the model and the general formal benchmark. In Sections \ref{sec:equilibrium} and \ref{sec:nonequilibrium}, we present the solution for the FPE in the equilibrium and non-equilibrium regimes, respectively. In Section \ref{sec:taylor} we revisit Taylor's law, showing a new macroecological pattern that goes beyond it. Section \ref{sec:speculation} then speculates on the physical origin of Taylor's law using an analogy with Brownian dynamics. Finally, in Section \ref{sec:discussion}, we discuss the general implications of our analysis and provide perspectives for future research, with a special focus on the consumer-resource model. 

\begin{figure*}[t]
    \centering
    \includegraphics[scale=0.5]{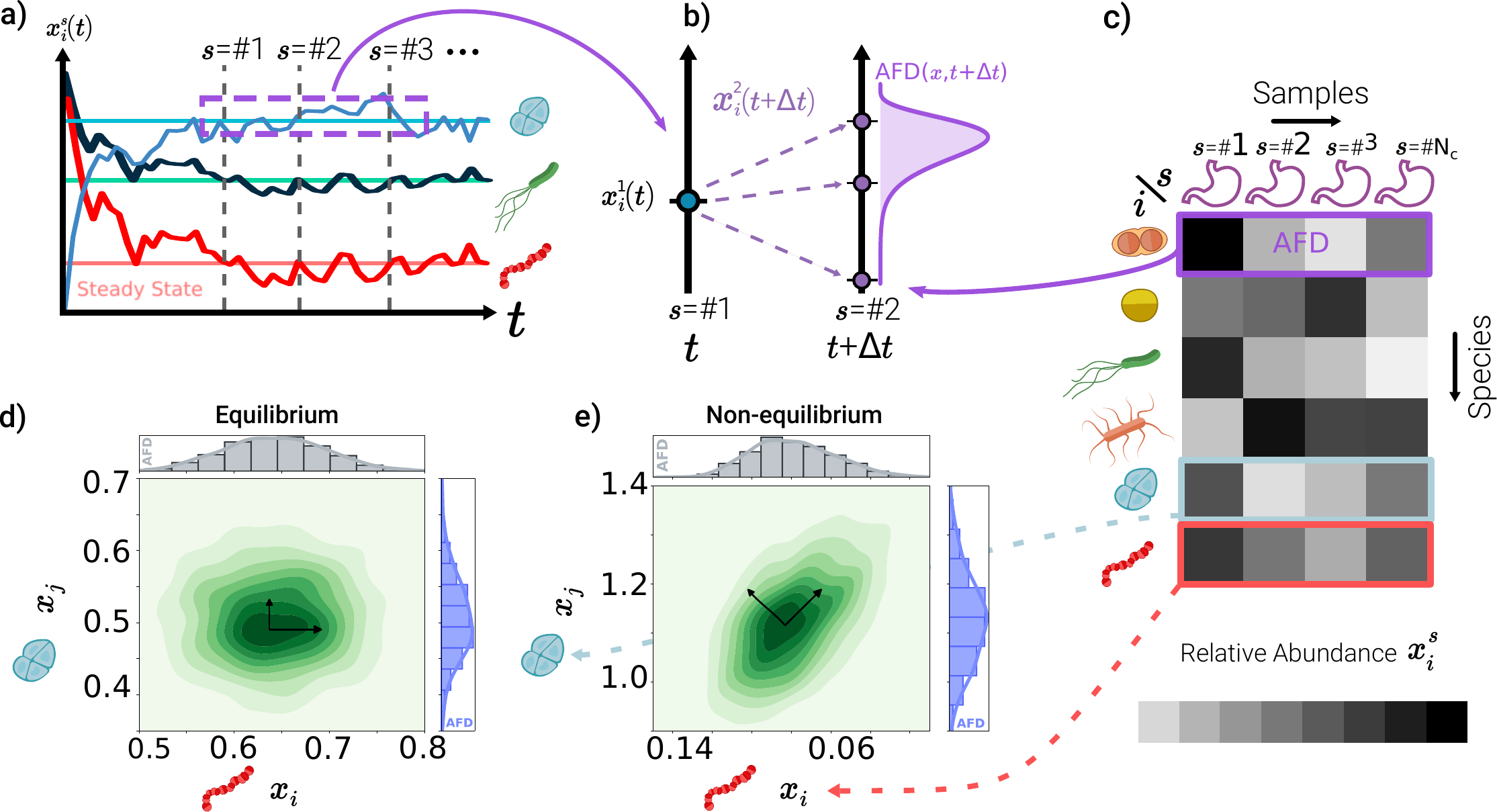}
    \caption{Infographic of the population dynamics and abundance correlation between two species.
    \textbf{a)} In-silico communities can be generated using the population model~\eqref{eq:mainEquation} with realistic parameter values. When the system reaches a stable state, the joint probability distribution $P(x_1,\dots,x_S)$ does not depend on time anymore.
    \textbf{b)} The fluctuations in the abundance of a given species are known to be distributed according to a gamma distribution. This Abundance Fluctuation Distribution (AFD) can be obtained as the marginal distribution $p_i(x_i)$ of the joint probability density.
    \textbf{c)} Information about species abundances in different samples may be arranged in a table for each biome, with species listed in rows and samples in columns. Herein, the AFD of species $i$ corresponds to the distribution of the elements in row $i$.
    \textbf{d)} In a state of equilibrium, where the system satisfies detailed balance, the joint probability distribution factorizes into a product of single-species distributions. As a result, there is no correlation between the abundances of different species.
    \textbf{e)} However, when detailed balance is not satisfied, the joint probability distribution no longer factorizes, and correlations between species abundances emerge.}
    \label{fig:1}
\end{figure*}

\section{Stochastic Lotka-Volterra model}
\label{sec:model}

\begin{figure*}
    \centering
    \includegraphics[scale=0.5]{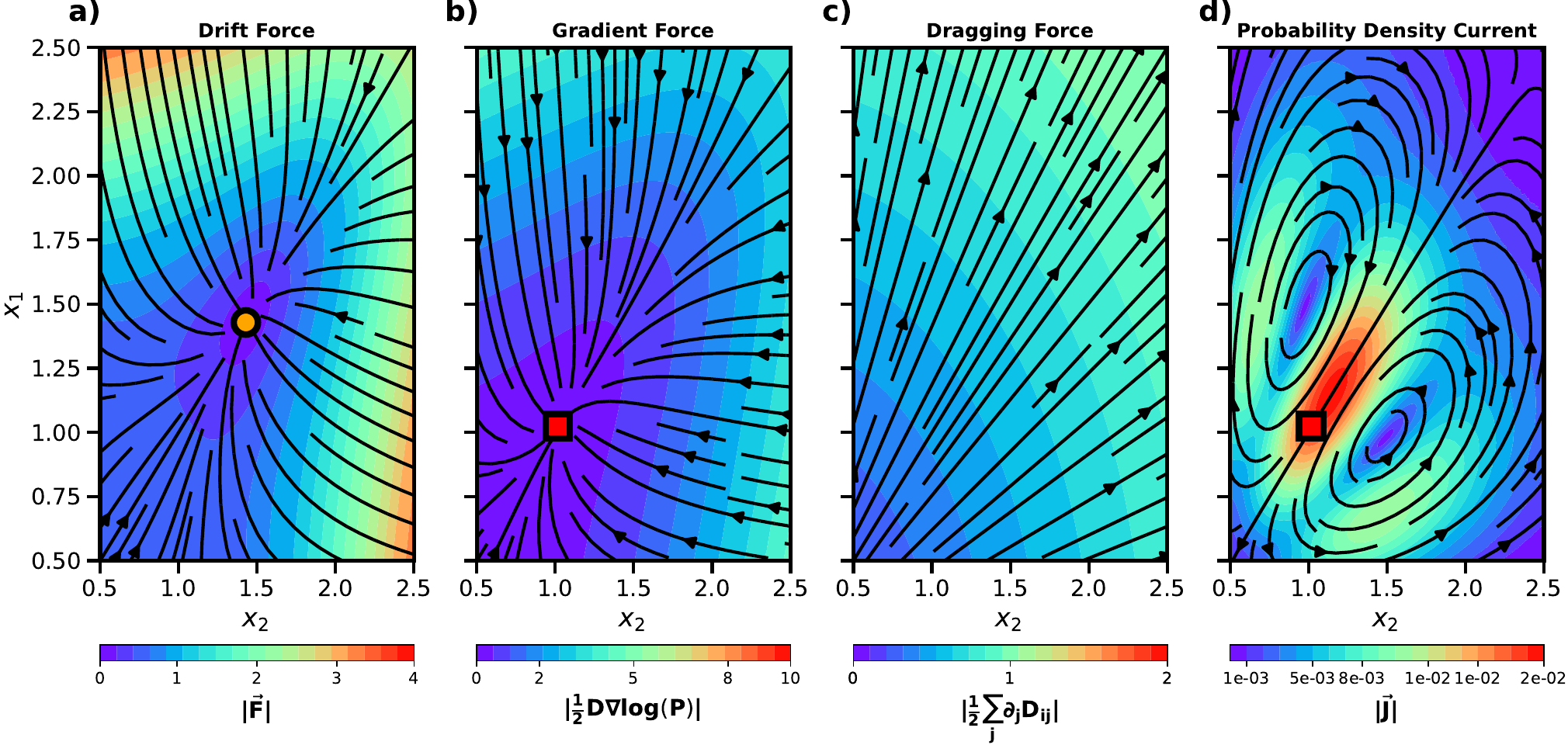}
    \caption{Force decomposition of non-equilibrium dynamics for a two-species ecosystem ($S=2$) in the stationary state. The time evolution of the probability of species abundances $P(\textbf{x},t)$ defined by Eq.~\eqref{eq:FPE} can be understood as the result of three forces: \textbf{a)} the drift force $F_i$, represented by the first term in Eq.~\eqref{eq:convectiveField}; \textbf{b)} the gradient force, corresponding to the second term in Eq.~\eqref{eq:convectiveField}; and \textbf{c)} the dragging force, related to the second term in the right hand-side of Eq.~\eqref{eq:current}. These three forces shape \textbf{d)} the probability density current (here we do not represent the curl-force for aesthetic visualization). Even though this decomposition of the gradient force is not unique, it is rather illustrative as the resulting dynamics turns out to be equivalent to that of charged particle moving within an electric and a magnetic field (the last one vanishing at equilibrium), and highlights the possibility of having closed loops in the current. The orange circle in \textbf{a)} marks the fixed point of the drift, and the red square in \textbf{b)} signals the maximum of the joint probability density. The data for the figure were obtained through a numerical integration of the FPE (see Appendix~\ref{app:Random-matrix-generation} for details). Parameters of the system: number of nodes in the square mesh $N_{x_1}=N_{x_2}=600$, length of the square mesh $L_{x_1}=L_{x_2}=6$, interaction matrix: $\mathbf{A}=\{a_{11}=a_{22}=-1, a_{12}=a_{21}=0.3\}$, noise matrix: $\mathbf{W}=\{w_{11}=w_{22}=0, w_{12}=w_{21}=0.3\}$, growth-rate $\tau_1=\tau_2=1$}.
    \label{fig:2}
\end{figure*}

A basic population model that includes both species (biotic) interactions and interrelated environmental (abiotic) fluctuations is the Stochastic Lotka-Volterra Model (SLVM) \cite{Camacho2024}, described by the system of equations (in It\^o's interpretation)
\begin{equation}\label{eq:mainEquation}
\begin{split}
    \dot{x}_i &=F_i(\vec{x})+x_i \xi_i(t), \\
    F_i(\vec{x}) &=\frac{x_i}{\tau_i} \left(1+\sum_{j=1}^S a_{ij} x_j\right),
\end{split}
\end{equation}
where $i=1,\dots,S$ runs over all $S$ species present in the community. Here, $x_i$ is the population (also termed abundance or density of individuals) of  species $i$; $\tau_i^{-1}$ its intrinsic growth rate; and $\xi_i(t)$ is a zero-mean, multivariate, Gaussian, white noise with correlations $\langle\xi_i(t)\xi_j(t^\prime)\rangle= w_{ij}\delta(t-t^\prime)$. Matrix $\mathbf{W}=(w_{ij})$ accounts for fluctuations of abiotic factors (e.g.~variation of nutrients, presence of chemicals, changes in temperature or pH, etc.) present in the environment. Along the same line, matrix $\mathbf{A}=(a_{ij})$ is an effective way to describe pairwise interactions between species, such as competition or cross-feeding. Its diagonal terms $a_{ii}=-1/K_i<0$ quantify intraspecific interactions through the carrying capacity of the species $K_i$, while the off-diagonal ones account for interspecific couplings and can have any sign. When the latter are zero, Eqs.~\eqref{eq:mainEquation} reduce to Grilli's Stochastic Logistic Model (SLM) \cite{grilli2020macroecological}---which we will also refer to as `non-interacting model'.

The SLVM \eqref{eq:mainEquation} rules the time evolution of the abundances of a community of $S$ species (see Fig.~\ref{fig:1}\textbf{a}) in the form of a system of Langevin equations. Alternatively, the same dynamics can be described through the FPE
\begin{equation}\label{eq:FPE}
\frac{\partial}{\partial t} P(\vec{x},t)+\nabla \cdot \vec{J}(\vec{x},t)=0,
\end{equation}
where $P(\vec{x},t)$ is the probability density of abundances $\textbf{x}$ at time $t$, and $\vec{J}(\vec{x},t)$ is the corresponding current, which can be expressed as
\begin{equation}\label{eq:current}
\vec{J}(\vec{x},t)=\vec{L}(\vec{x})P(\vec{x},t)
-\frac{1}{2}\vec{D}(\vec{x})\nabla P(\vec{x},t),
\end{equation}
with the functions 
\begin{equation}\label{eq:diffusionMat}
D_{ij}(\vec{x})=x_ix_jw_{ij}
\end{equation}
playing the role of a diffusion matrix term in the FPE, and
\begin{equation}
\begin{split}\label{eq:convectiveField}
L_i(\vec{x}) &=F_i(\vec{x})-
\frac{1}{2}\sum_{j=1}^S\frac{\partial D_{ij}}{\partial x_j} \\
&=\frac{x_i}{\tau_i}\left[1-\frac{\tau_iw_{ii}}{2}
+\sum_{j=1}^S\left(a_{ij}x_j
-\frac{\tau_iw_{ij}}{2}\right)\right].
\end{split}
\end{equation}
exhibiting the form of a convective field \cite{mendler2020predicting}.

The basic phenomenology of the microbiome observed in real systems involves rapid random fluctuations of abundances around a stable state \cite{gibbons2017two}. In the broader context of macroecology, a natural approach to characterize these fluctuations is by examining their distribution along different time instants or, more generally, across different samples. This distribution is referred to as the Abundance Fluctuation Distribution (AFD) (Fig.~\ref{fig:1}\textbf{b}), and it has been reported to be well fitted by a gamma distribution regardless of the ecosystem \cite{grilli2020macroecological}. 

In terms of the solution of the FPE, the AFD corresponds to the marginal distribution $p_i(x_i)$ of the probability density in the steady state. The latter is obtained by solving the equation $\nabla\cdot\vec{J}(\vec{x})=0$. This equation admits a large display of solutions though. The simplest one is $\vec{J}(\vec{x})=\vec{0}$, which corresponds to a system \emph{in equilibrium}. Accordingly, solutions $\vec{J}(\vec{x})\ne\vec{0}$ describe stationary states that are \emph{out of equilibrium}. We shall discuss both cases separately.

\section{Equilibrium dynamics}
\label{sec:equilibrium}

One obtains the probability distribution in equilibrium by solving the equation $\vec{J}(\vec{x})=\vec{0}$. Using expression \eqref{eq:current} for the current, this leads to
\begin{equation} \label{eq:gradient-eq-vec}
\nabla\log P(\vec{x})=2\mathbf{D}(\vec{x})^{-1}\vec{L}(\vec{x}).
\end{equation}
Recalling the expressions for the diffusion matrix \eqref{eq:diffusionMat} and the convective field \eqref{eq:convectiveField}, we end up with the set of equations
\begin{equation}
\label{eq:gradient-eq}
\frac{\partial\log P}{\partial x_i} =\frac{1}{x_i}\left(\beta_i-1
+2\sum_{k,j=1}^S\frac{m_{ik}a_{kj}x_j}{\tau_k}\right), 
\end{equation}
where we have introduced
\begin{equation}\label{eq:betas}
\beta_i \equiv \sum_{k=1}^Sm_{ik}\left(\frac{2}{\tau_k}-w_{kk}\right),
\end{equation}
$m_{ij}$ being the entries of $\mathbf{M}=\mathbf{W}^{-1}$. To allow for a solution, the aforementioned set of equations must satisfy a detailed balance condition \cite{gardiner2004} (derived in Appendix~\ref{app:FPEequilibrium}), which links the interaction and noise matrices $\vec{A}$ and $\vec{W}$ [c.f.~Eq.~\eqref{eq:gradient-matrix}]. If this condition holds, the stationary distribution of the microbial community takes the form
\begin{align}
\label{eq:factorizedSolution}
P(\vec{x}) &=\prod^S_{i=1}p_i (x_i), \\
\label{eq:gammaSolution}
p_i(x_i) &=\frac{1}{\bar{x}_i\Gamma(\beta_i)}
\left(\frac{x_i}{\bar{x}_i}\right)^{\beta_i-1}
e^{-\beta_ix_i/\bar{x}_i},  
\end{align}
where $p_i$ is the AFD of species $i$. The form of this distribution is the same as that obtained without species interactions and a diagonal noise matrix $\vec{W}$ \cite{grilli2020macroecological}, however the shape parameters $\beta_i$ and the average abundances $\bar{x}_i$ are different because their values also depend on the interactions.

Nonetheless, the most significant feature of the aforementioned result is that the community distribution \eqref{eq:factorizedSolution} factorizes into distinct components, each associated with a different species. This implies that the abundances of individual species are uncorrelated and fluctuate independently of each other (see Figs.~\ref{fig:2}\textbf{c} and \textbf{d}), in contrast to what is observed in empirical data. The latter show non-trivial correlations often spanning the entire range of Pearson's values \cite{Camacho2024,grilli2020macroecological,Ho2022} (see Fig.~\ref{fig:2}\textbf{e}). They are the fingerprint of an inherent non-equilibrium microbial dynamics.

\subsection{The emergence of correlations in Consumer-Resource models}\label{sec:CR}

While the equilibrium solution cannot describe real communities due to its inability to predict correlations, it nevertheless sheds light on ecological mechanisms beyond the model we are considering. In particular, it is well-known in theoretical ecology that consumer-resource (CR) models can be cast into Lotka-Volterra form. The former explicitly account for the dynamics of species abundances as well as their resource consumption, leading to interactions driven by the limitation of resources and the competition between species occupying similar niches. 

Recently, some CR models have been introduced to analyze the correlations between populations of different species in terms of their phylogenetic distance \cite{sireci2023environmental}. In them, the relationship between resources and consumers is represented through two matrices, $\textbf{B}$ and $\textbf{Q}$. The preference of species $i$ for resource $r(=1,\dots,R)$ is described by the $i^{\rm th}$ column of the $R\times S$ matrix $\textbf{B}$. Likewise, the columns of matrix $\textbf{Q}$ encode the preference of species for certain environmental conditions. These preference vectors have only positive entries and are all normalized to the same constant value. In the mapping to a Lotka-Volterra model, matrices $\textbf{W}$ and $\textbf{A}$ are obtained from $\textbf{B}$ and $\textbf{Q}$ as
\begin{equation}
\mathbf{W} = \omega \mathbf{B}^{\rm T}\mathbf{B}
+\nu\mathbf{Q}^{\rm T}\mathbf{Q}, \qquad
\mathbf{A} = -\gamma \mathbf{B}^{\rm T}\mathbf{B},
\end{equation}
where the coefficient $\omega$, $\nu$, and $\gamma$ weight the effect of the different ecological forces. 

In this framework, the authors of \cite{sireci2023environmental} show that, in cases where the evolution of the system is dominated by competition for fluctuating shared resources ($\nu=0$), correlations between abundances are absent. Note that this regime corresponds to a situation where $\mathbf{A}$ and $\mathbf{W}$ are proportional, defining an equilibrium state characterized by the balance between abiotic and biotic forces. Specifically, under the approximation in which the intrinsic growth rates are the same for all the species ($\mathbf{T}=\tau\mathbf{I}$), 
\begin{equation}
\mathbf{W}=-\frac{\gamma}{\omega}\mathbf{A}.
\end{equation}
This equation is equivalent to the equilibrium condition \eqref{eq:gradient-matrix} if the choice $\mathbf{E}=-(\gamma/\omega\tau)\mathbf{I}$ (see Appendix~\ref{app:FPEequilibrium}) is made in that equation. It is thus not surprising that a system described by this particular CR model, which is at equilibrium according to our previous analysis, lacks correlations between species, as evidenced by the numerical simulations of Ref.~\cite{sireci2023environmental}.

\section{Non-equilibrium dynamics}\label{sec:nonequilibrium}

The lack of correlations of the equilibrium solution, in contrast with its presence in real microbiomes, prompts an exploration of the dynamics out of equilibrium. The FPE has a stationary nonequilibrium solution if Eq.~\eqref{eq:current} can be solved for some $\vec{J}(\vec{x})\ne\vec{0}$ such that $\nabla\cdot\vec{J}=0$.

The general solution of $\nabla\cdot\vec{J}=0$ is \cite{flanders1963differential}
\begin{equation}\label{eq:rotational}
J_i(\vec{x})=\sum_{j_2,\dots,j_S=1}^S\epsilon_{ij_2\dots j_S}
\frac{\partial G_{j_3\dots j_S}(\vec{x})}{\partial x_{j_2}},
\end{equation}
where $\epsilon_{i_1\dots i_S}$ is the Levi-Civita symbol ($=1$/$-1$ if $\{i_1,\dots,i_S\}$ is an even/odd permutation of $\{1,\dots,S\}$, and $=0$ otherwise) and $G_{j_3\dots j_S}(\vec{x})$ is any fully anti-symmetric tensor. Accordingly, there are $S(S-1)/2$ unknown functions, to be determined under the assumption that Eq.~\eqref{eq:current} can be solved for $P(\vec{x})$.

In general, the study of non-equilibrium dynamics is a rather complicated problem that has been solved only in specific cases \cite{risken1996fokker}. Physically, the complexity arises from the fact that Eq.~\eqref{eq:current} reflects the interplay between three types of forces---whose components are schematically depicted in Fig.~\ref{fig:2} \cite{wang2015landscape}. Whereas at equilibrium detailed balance renders a null current, out of equilibrium the stationary state has cycles (see Fig.~\ref{fig:2}\textbf{d}).


But we can be less ambitious and try instead to determine the marginal distribution $p_i(x_i)$ for the abundance of species $i$. An equation for it can be obtained by integrating out in the FPE all degrees of freedom except $x_i$. The resulting equation is (see Appendix~\ref{app:marginal} for its derivation)
\begin{equation}\label{eq:marginalized-FP}
0=-\frac{\partial}{\partial x_i}
\big[f_i(x_i)p_i(x_i)\big]+\frac{w_{ii}}{2}
\frac{\partial^2}{\partial x_i^2}\big[x_i^2p_i(x_i)\big],
\end{equation}
where $f_i(x_i)$ is given by Eq.~\eqref{eq:drift-i}. This equation admits the solution (up to normalization)
\begin{equation}\label{eq:general-marginal-xk}
\begin{split}
p_i(x_i)&\propto\, x_i^{\frac{2}{\tau_iw_{ii}}-2}
\exp\left(-\frac{2x_i}{\tau_iw_{ii}K_i}\right) \\ 
&\times\exp\left(\frac{2}{\tau_iw_{ii}}\sum_{j\ne i}^Sa_{ij} 
\int\bar{x}_j(x_i)\frac{dx_i}{x_i}\right).
\end{split}
\end{equation}
In the absence of interactions, the last exponential is $1$ and we recover the solution of the SLM \cite{grilli2020macroecological}. The extra factor introduced by interactions involves the calculation of the conditional average $\bar{x}_j(x_i)$, defined in Eq.~\eqref{eq:cond-av}.

Equation~\eqref{eq:general-marginal-xk} is as far as we can get in computing the marginal probability density because there is no easy way to compute $\bar{x}_j(x_i)$ other than solving the FPE itself. However, we can use empirical information to figure out a plausible approximation for this function.

\subsection{An approximation for the marginal probability density}
\label{sec:ansatz}

Numerical simulations performed on the SLVM~\eqref{eq:mainEquation} with nontrivial interaction matrices \cite{Camacho2024}, show that the resulting AFD can still be well-fitted by a gamma distribution. The only way that such a distribution can be obtained from \eqref{eq:general-marginal-xk} is if
\begin{equation}\label{eq:ansatz-hypothesis}
\bar{x}_j(x_i)=\eta_{ji}+\varphi_{ji}x_i,
\end{equation}
with $\eta_{ji}$ and $\varphi_{ji}$ two as yet undetermined sets constants. Most likely, the gamma shape of $p_i(x_i)$ is only an approximate result, so the dependence \eqref{eq:ansatz-hypothesis} should be taken as an educated ansatz.

\begin{figure}
    \centering
    \includegraphics[scale=0.36]{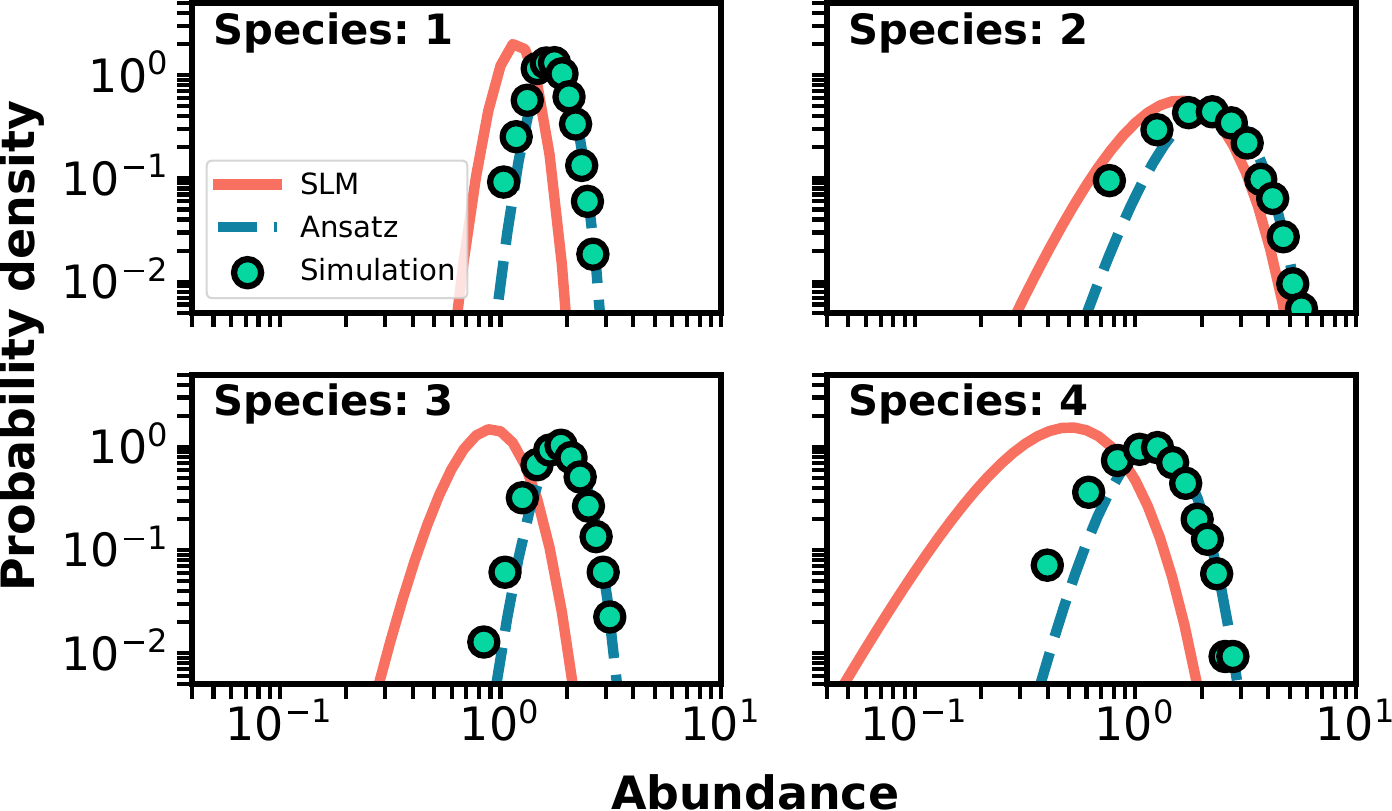}
    \caption{\label{fig:3} The figure portrays the AFDs of an \textit{in-silico} interacting (out of equilibrium) community with $S=4$ species. Green dots depict the distributions as obtained from simulations of the SLVM, Eqs.~\eqref{eq:mainEquation}. Blue dashed lines corresponds to the gamma distributions obtained by solving numerically Eqs.~\eqref{eq:implicit-mean} and \eqref{eq:LV-solution-linear-system}. Red lines depict the AFDs for the same community, but without interactions. The effect that interactions have on the mean and variance of the AFDs is striking. Simulations were carried out as specified in Appendix~\ref{app:Random-matrix-generation}, with parameters $\tau=0.1$, $C=1$, $\mu_l=0.1, \sigma_l=0.5$, $\sigma_n=0.2$, $\Delta t=10^{-3}$.}
\end{figure}

\begin{figure*}
\centering
   \includegraphics[width=1\textwidth]{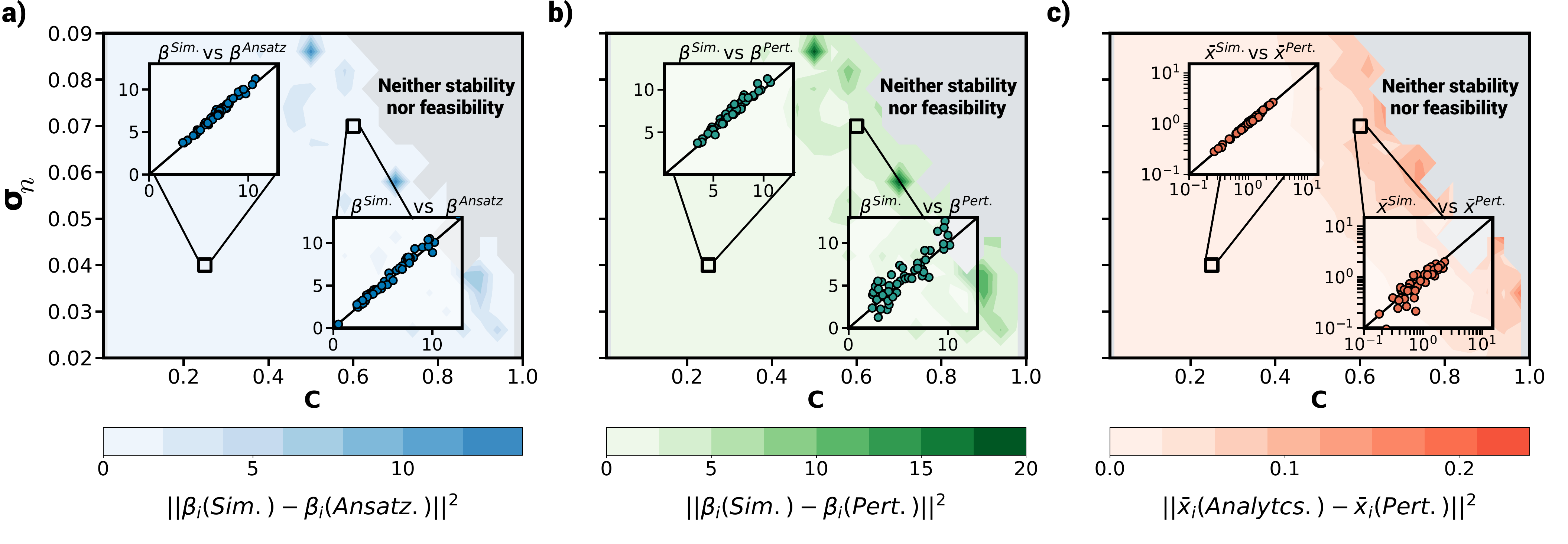}
   \caption{Accuracy of the ansatz and the perturbative expansion \eqref{eq:perturb-expansions}--\eqref{eq:pert-beta-1} across the set of parameters that characterize the species interaction matrix $\mathbf{A}$. The first two panel compare the shape parameters ($\beta_i$) obtained from simulations of the Langevin equations with those obtained using \textbf{a)} the ansatz and \textbf{b)} the perturbative expansion. Panel \textbf{c)} compares the mean abundances obtained from the perturbative expansion with the exact ones [computed through Eq.~\eqref{eq:LV-solution-linear-system}]. Each pixel represents a unique combination of network connectance $C$ (horizontal axis) and standard deviation $\sigma_n$ of the interaction strength distribution (vertical axis), with the pixel color indicating the mean squared error (averaged over $100$ realizations) between the reference values and the approximate predictions. Gray regions denote parameter spaces where systems are neither stable nor feasible. Insets depict examples of realizations, with the black line representing the identity line $y=x$. For this figure, we set $\tau=0.1$, $S=50$, $\mu_l=0.1$, $\sigma_l=0.5$, and $\Delta t=10^{-3}$. A non-trivial noise matrix $\textbf{W}$ is generated for each realization as described in Appendix \ref{app:Random-matrix-generation}.}
   \label{fig:4}    
\end{figure*}


Several consistency conditions will help us determine the parameters $\eta_{ji}$ and $\varphi_{ji}$. First, by averaging $\bar{x}_j(x_i)$ we obtain
\begin{equation}\label{eq:mean-consistency}
\bar{x}_j = \int_0^{\infty}p_i(x_i)\bar{x}_j(x_i)\,dx_i= \eta_{ji}+\varphi_{ji}\bar{x}_i,
\end{equation}
which in particular implies $\eta_{ii}=0$ and $\varphi_{ii}=1$. Likewise, we can obtain second moments as
\begin{equation}\label{eq:correlator-consistency}
\overline{x_jx_i}=\int_0^{\infty}\bar{x}_j(x_i)x_ip_i(x_i)\,dx_i= \eta_{ji}\bar{x}_i+\varphi_{ji}\overline{x^2_i}.
\end{equation}
Solving the system~\eqref{eq:mean-consistency}, \eqref{eq:correlator-consistency} yields
\begin{equation}
\varphi_{ji}=\frac{\Cov(x_j,x_i)}{\sigma_i^2}=\frac{\sigma_j}{\sigma_i}\rho_{ji},
\qquad \eta_{ji}=\bar{x}_j-\varphi_{ji}\bar{x}_i,
\end{equation}
where $\rho_{ji}$ are Pearson's correlation coefficients and $\sigma_i$ is the standard deviation of $p_i(x_i)$. Rewriting everything in terms of the shape parameters $\beta_i=\bar{x}_i^2/\sigma_i^2$ of the gamma distribution [c.f.~Eq.~\eqref{eq:gammaSolution}], we finally obtain
\begin{equation}\label{eq:cond-mean-pears}
\bar{x}_j(x_i)=\bar{x}_j\left[1+\rho_{ji}\sqrt{\frac{\beta_i}{\beta_j}} 
\left(\frac{x_i}{\bar{x}_i}-1\right)\right].
\end{equation}

Equation~\eqref{eq:general-marginal-xk} with the conditional average \eqref{eq:cond-mean-pears} becomes a gamma distribution. Identifying its expression with that of Eq.~\eqref{eq:gammaSolution} leads to the equations
\begin{align}\label{eq:implicit-mean}
&\beta_i+\frac{2}{\tau_iw_{ii}}\sum_{j=1}^Sa_{ij}
\bar{x}_j\rho_{ji}\sqrt{\frac{\beta_i}{\beta_j}}=0, \\
\label{eq:LV-solution-linear-system}
&1-\frac{\tau_iw_{ii}}{2}+\sum_{j=1}^Sa_{ij}\bar{x}_j=0,
\end{align}
where we have made use of the identities $a_{ii}=-1/K_i$, $\rho_{ii}=1$. It is worth noticing that Eq.~\eqref{eq:LV-solution-linear-system} is an exact equation that can be derived directly from the Langevin Eqs.~\eqref{eq:mainEquation} under the assumption that the system is in a stationary state \cite{Camacho2024}.

Equation~\eqref{eq:LV-solution-linear-system} is a linear system that can be numerically solved for the averages $\bar{x}_i$ using standard methods. As for Eq.~\eqref{eq:implicit-mean}, solving it requires knowledge of the Pearson's correlation coefficients $\rho_{ij}$. By using the values of $\rho_{ij}$ obtained from simulations, we can solve Eq.~\eqref{eq:implicit-mean} for the shape parameters $\beta_i$ through a Levenberg–Marquardt algorithm \cite{Levenberg1944, Marquardt1963}. In Fig.~\ref{fig:3}\textbf{a} we compare the AFD obtained with this ansatz with numerical simulations of Eqs.~\eqref{eq:mainEquation} for four different species. The good agreement reached in all cases validates the ansatz \eqref{eq:cond-mean-pears}. For reference, we also plot the AFD of the non-interacting system ($a_{ij}=-\delta_{ij}/K_i$). Notably, even though the distributions with and without interactions have both the shape of a gamma, their mean values and variances differ significantly.

But perhaps the most remarkable result of this approximation is that the shape parameters $\beta_i$ explicitly depend on species interactions, which means that they must necessarily have a nontrivial distribution, even if the products $w_{ii}\tau_i$ were constant for all species. As interactions are unavoidable to explain correlations \cite{Camacho2024}, the existence of this variability becomes a prediction of the present theory which we will explore in more depth in Sec.~\ref{sec:taylor}.

In Fig.~\ref{fig:4}\textbf{a} we explore the limits of validity of the proposed ansatz across the set of parameters that characterize the species interaction matrix $\mathbf{A}$. The analysis reveals that, when interactions are generated according to a zero-mean, normal distribution, the ansatz proves highly accurate throughout the entire region of feasibility and stability.

\subsection{Perturbative expansion}\label{sec:perturbative}


As we have just seen, Eqs.~\eqref{eq:LV-solution-linear-system} and \eqref{eq:implicit-mean} can be numerically solved for $\bar{x}_i$ and $\beta_i$, but we need Pearson's matrix $\rho_{ij}$ as an input. This problem can be circumvented though, provided we assume that the off-diagonal elements of the interaction matrix $\mathbf{A}$ are ``small'', because then we can try a perturbative approach to solve those equations.  (Small in this context means $\sum_{j\ne i}K_i|a_{ij}|\ll 1$ for all $i$.) In what follows we will assume that the environmental noise matrix $\mathbf{W}$ is diagonal---which means that $\rho_{ij}=\delta_{ij}$ in the non-interacting limit.

Let us assume that $\bar{x}_i$ and $\beta_i$ have the expansions
\begin{equation}\label{eq:perturb-expansions}
\bar{x}_i=\bar{x}_i^{(0)}+\bar{x}_i^{(1)}+\cdots, \quad
\beta_i=\beta_i^{(0)}+\beta_i^{(1)}+\cdots,
\end{equation}
where the terms with superscript $n$($=0,1,\dots$) are assumed to be of order $O(a_{ij}^n)$ ($i\ne j$). Then, Eq.~\eqref{eq:LV-solution-linear-system} becomes
\begin{equation*}
1-\frac{\tau_iw_{ii}}{2}-\frac{\bar{x}_i^{(0)}}{K_i}-\frac{\bar{x}_i^{(1)}}{K_i}-\cdots
+\sum_{j\ne i}^Sa_{ij}\bar{x}_i^{(0)}+\cdots=0,
\end{equation*}
from which, comparing like orders,
\begin{align}
&\bar{x}_i^{(0)}=K_i\left(1-\frac{\tau_iw_{ii}}{2}\right), \label{eq:pert-mean-0} \\
&\bar{x}_i^{(1)}=K_i\sum_{j\ne i}^Sa_{ij}\bar{x}_j^{(0)}. \label{eq:pert-mean-1}
\end{align}
Likewise, since $\rho_{ij}=\delta_{ij}+\omega_{ij}$, with $\omega_{ii}=0$ and $\omega_{ij}=O(a_{ij})$ if $i\ne j$, Eq.~\eqref{eq:implicit-mean} becomes
\begin{equation*}
\beta_i^{(0)}+\beta_i^{(1)}+\cdots=
\frac{2(\bar{x}_i^{(0)}+\bar{x}_i^{(1)}+\cdots)}{\tau_iw_{ii}K_i}+\cdots.
\end{equation*}
Comparing orders,
\begin{align}
&\beta_i^{(0)}=\frac{2}{\tau_iw_{ii}}-1, \label{eq:pert-beta-0} \\
&\beta_i^{(1)}=\frac{2}{\tau_iw_{ii}}\sum_{j\ne i}^Sa_{ij}\bar{x}_j^{(0)}.
\label{eq:pert-beta-1}
\end{align}
Thus, while calculating higher orders in expansions \eqref{eq:perturb-expansions} would require the knowledge of Pearson's matrix, a first-order approximation can be obtained without it.

As illustrated by Figs.~\ref{fig:4}\textbf{b} and \textbf{c}, the perturbative approximation proves to be sufficiently accurate across most of the feasibility and stability region, when interaction matrices are randomly generated from a normal distribution N$(0,\sigma_n)$. Discrepancies only show up near the boundary of this region, specifically at the frontier of the stability-feasibility domain. Notably, the bottom inset in Fig.~\ref{fig:4}\textbf{b} highlights a significant deviation between the simulated values of the shape parameters and those predicted by our perturbative expansion. This discrepancy is mainly due to the high values of connectance ($C$) and interaction strength ($\sigma$), which cannot be fully grasped within a perturbative framework.

\begin{figure}
    \centering
    \includegraphics[width=0.48\textwidth]{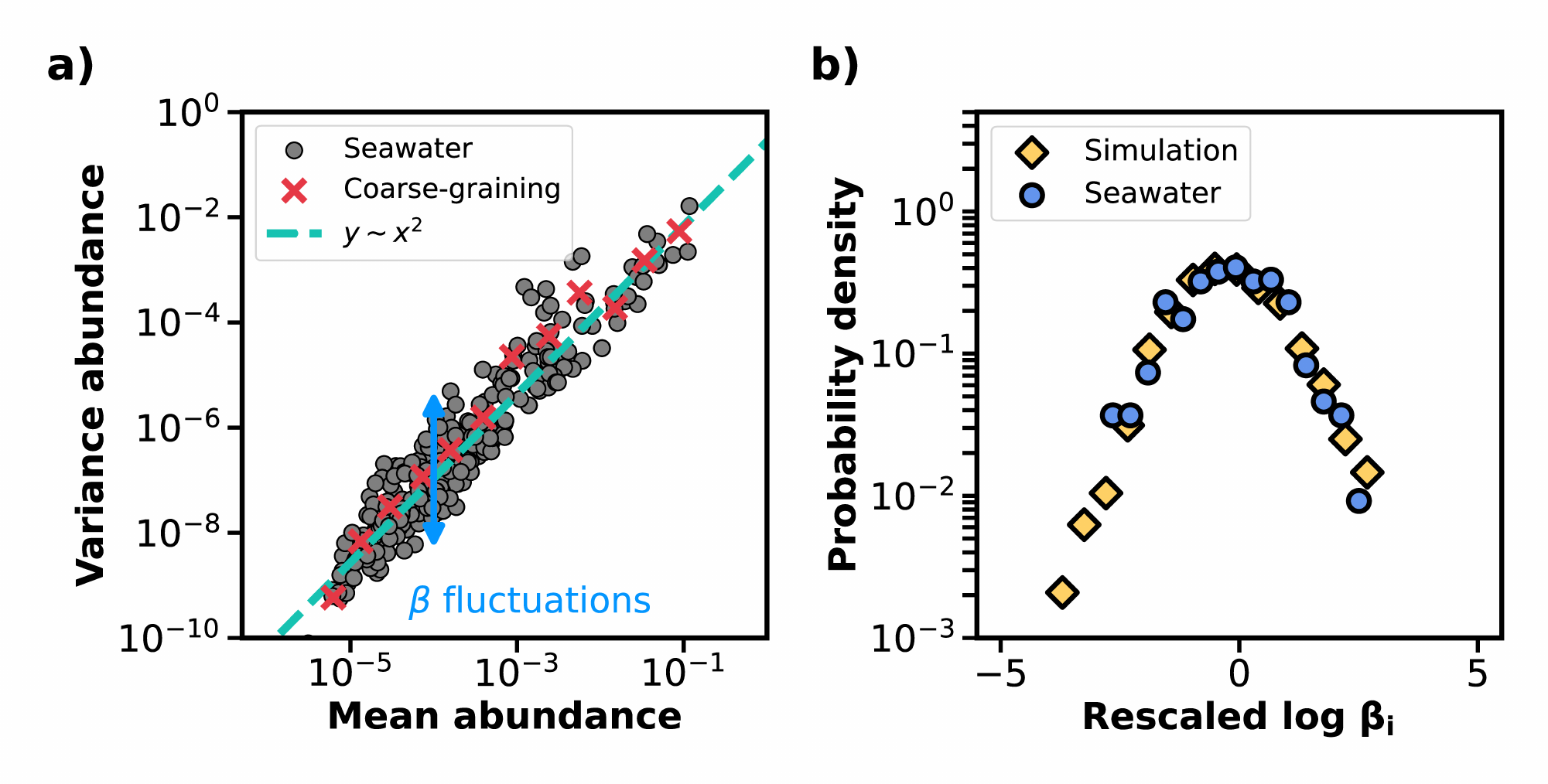}
    \caption{\label{fig:5} Distribution of shape parameters $\beta_i$ as obtained from empirical data of the EBI Metagenomics \cite{Mitchell2018} Seawater microbiome, as well as from numerical simulations of the SLVM \eqref{eq:mainEquation}.
    \textbf{a)} Log-log plot of the empirical data for variance $\sigma_i^2$ vs.~mean $\bar{x}_i$ of the species abundances (grey dots). Red crosses are the result of partitioning the whole interval of mean abundances (in logarithmic scale) and averaging these data within bins. The dashed line represents the curve $\sigma^2=\beta \bar{x}^2$, corresponding to Taylor's law. While globally this curve is an excellent fit to the data, the vertical variability (indicated by a double blue arrow) spans several orders of magnitude. This variability is masked in the coarse-grained data (red crosses).
    \textbf{b)} Probability density of the re-scaled log-shape parameter [c.f.~Eq.~\eqref{eq:rescaled-logshape}] as obtained from the empirical data (blue dots) of the Seawater microbiome, as well as from simulations of the SLVM (yellow diamonds).
    In making this figure, species appearing in less than $30\%$ of the samples were filtered out. Simulations were carried out as specified in Appendix~\ref{app:Random-matrix-generation}, with parameters $S=300,\tau=0.1$, $C=0.5$, $\mu_l=0, \sigma_l=1.5$, $\sigma_n=10^{-3}$, $\Delta t=10^{-3}$.}
\end{figure}

One of the most noteworthy consequences of this perturbative expansion is that it explicitly reveals that species interactions alone can explain the variability of the shape factor [c.f.~Eq.\eqref{eq:pert-beta-1}].
This variability is also present in the empirical data, as Fig.~\ref{fig:5} illustrates. Fig.~\ref{fig:5}\textbf{b} also shows that the results of simulations of the SLVM closely replicate the empirical distributions of shape parameters.

\subsection{Physical interpretation}

A mean-field (MF) approach to the Langevin equations \eqref{eq:mainEquation} can shed light on the meaning of the ansatz proposed in Sec.~\ref{sec:ansatz}. When the system is in a steady state, we can approximate the driving force $F_i(\vec{x})$ by
\begin{equation}\label{eq:mean-field-drift}
\begin{split}
F^{\rm MF}_i(x_i,\vec{\bar{x}}) &=\frac{x_i}{\tau_i}
\left(q_i(\vec{\bar{x}})-\frac{x_i}{K_i}\right), \\
q_i(\vec{\bar{x}}) &\equiv 1+\sum_{j \neq i}^Sa_{ij}\bar{x}_j.
\end{split}
\end{equation}
The rationale is that the effect of the abundances of other species on species $i$ is an average of quantities that fluctuate around some mean values, and this averaging smooths out fluctuations. In this approximation, the SLVM becomes a SLM with effective growth rates and carrying capacities
\begin{equation}\label{eq:mean-field-K-tau}
\tau_i^*(\vec{\bar{x}})\equiv\frac{\tau_i}{q_i(\vec{\bar{x}})},
\qquad K^*_i(\vec{\bar{x}})\equiv K_iq_i(\vec{\bar{x}}).
\end{equation}
Thus, the solution of the FPE is a product of gamma distributions with parameters
\begin{align}
\bar{x}_i &=\left(1-\frac{\tau_i^*(\vec{\bar{x}})w_{ii}}{2}\right)
K^*_i(\vec{\bar{x}}),
\label{eq:mean-beta-MF-as-SLM} \\
\beta_i &=\frac{2}{w_{ii}\tau_i^*(\vec{\bar{x}})}-1.
\label{eq:MF-beta-final}
\end{align}
Remarkably, Eq.~\eqref{eq:mean-beta-MF-as-SLM} is identical to \eqref{eq:LV-solution-linear-system}, while at the same time \eqref{eq:MF-beta-final} can be obtained from \eqref{eq:implicit-mean} by replacing $\rho_{ij}=\delta_{ij}$ (no correlations). Interestingly, to first order in the interaction constants these two equations lead to the same perturbative expansions \eqref{eq:pert-mean-0}--\eqref{eq:pert-beta-1}.

The conclusion from this analysis is that the ansatz coincides with a MF approximation to first order in the expansion on the interaction coefficients, but beyond that it provides a refinement on MF by including nontrivial correlations in the equation for the shape parameters \eqref{eq:implicit-mean}.


\section{Taylor's law revisited: A new macroecological pattern}\label{sec:taylor}

In ecology, Taylor's law \cite{Taylor1961} describes a power-law relationship between the mean and variance of species abundances, namely
\begin{equation}\label{eq:TaylorLaw}
\sigma_i^2 = b\bar{x}_i^{a}, \quad a,b>0.
\end{equation}
With exponent $a=2$, it has been identified as a macroecological pattern in microbiomes \cite{grilli2020macroecological} (also known as Grilli's second law \cite{Camacho2024}). A consequence of this pattern is that the shape factors $\beta_i$ of the gamma AFDs for the different species in a microbiome are all the same ($\beta_i=b^{-1}$). However, in the light of the results that we have obtained in the previous sections (c.f.~Fig.~\ref{fig:5}), this macroecological pattern must be nuanced.

\begin{figure*}
    \centering
    \includegraphics[scale=0.27]{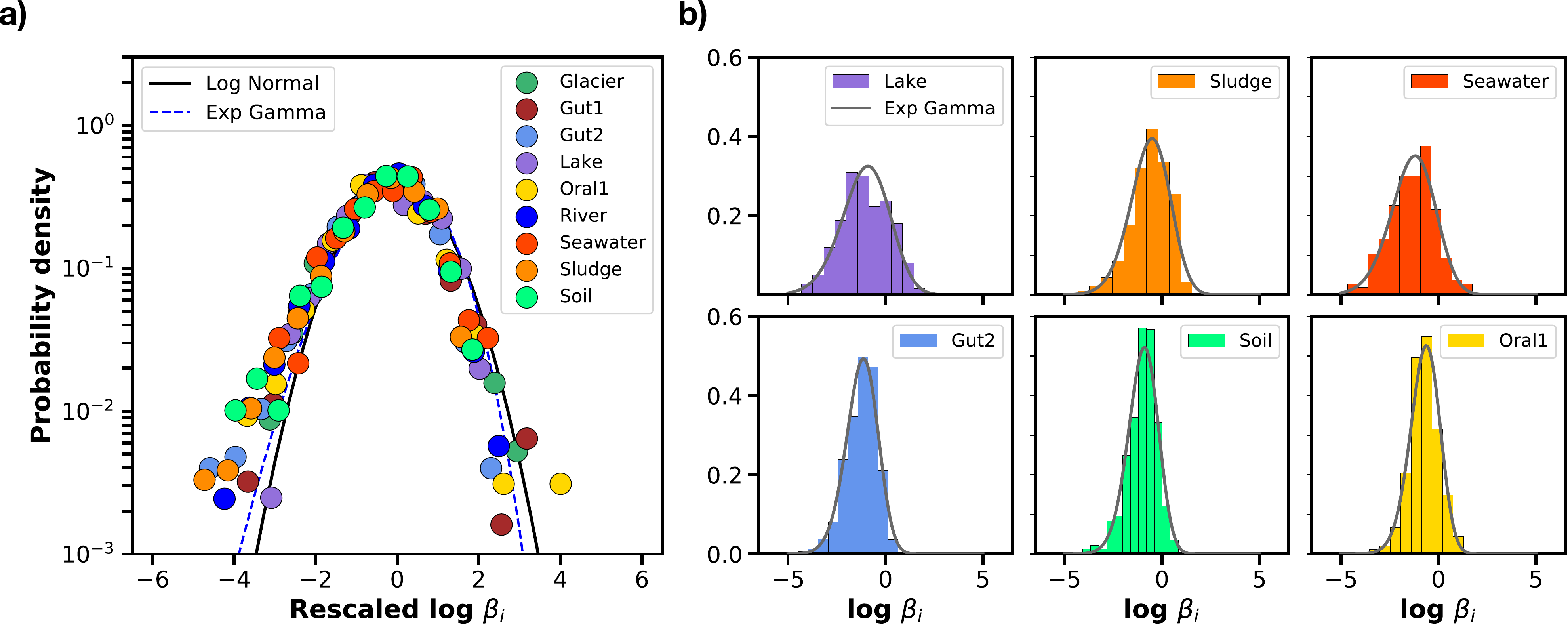}
    \caption{\label{fig:6} \textbf{a)} Probability density of the re-scaled log-shape parameter $z$ defined in Eq.~\eqref{eq:rescaled-logshape}, for all the species in several datasets of the EBI Metagenomics platform \cite{Mitchell2018}. The solid line represents the standard normal distribution $\log P=-z^2/2-\log\sqrt{2\pi}$ (likelihood: 16721); the dashed line is a best fit (with $\alpha=14$) to a standard exp-gamma distribution $\log P=\log[c/\Gamma(\alpha)]+\alpha(a+cz)-e^{a+cz}$, with $a=\psi(\alpha)\equiv\Gamma'(\alpha)/\Gamma(\alpha)$ the digamma function \cite[\S 5.2(i)]{NIST:DLMF} and $c=\sqrt{\psi'(\alpha)}$ (likelihood: 38095). \textbf{b)} Probability density of the log-shape parameters without re-scaling for several microbiomes, along with the specialization of the exp-gamma distribution to their specific means and standard deviations (grey curves). This representation makes clear the actual width of the empirical distributions as well as the goodness of the fit to an exp-gamma distribution.
    }
\end{figure*}

To begin with, Fig.~\ref{fig:6} plots the values of the rescaled log-shape parameter, defined as
\begin{equation}\label{eq:rescaled-logshape}
z=\frac{\log\beta-\overline{\log\beta}}{\sqrt{\Var(\log\beta)}},
\end{equation}
for all the species in several datasets of the EBI Metagenomics platform \cite{Mitchell2018}. Not only does this figure show the variability in the shape parameters, but it also shows that the data for all microbiomes collapse in the same universal curve. This provides a novel macroecological pattern. Fig.~\ref{fig:6} depicts two tentative fits to the data: log-normal and exp-gamma \footnote{The distribution followed by the logarithm of a gamma-distributed variable.} distributions. Not only does the fit to an exp-gamma has a much higher likelihood than the fit to a log-normal, but it also captures the evident skewness of the distributions (see Fig.~\ref{fig:6}\textbf{b}).

Even though we lack a theoretical underpinning that supports any specific shape for what this distribution should be, Grilli's third law provides empirical evidence for a log-normal distribution of mean abundances \cite{grilli2020macroecological}. As the variability in the distribution of $\beta_i$ arises from a linear combination of the mean abundances---as evidenced in the first order term \eqref{eq:pert-beta-1} of the perturbative expansion---this variability is expected to behave as a linear combination of log-normal random variables. Although such a combination does not have a known analytic form, it is nevertheless consistent with a gamma or log-normal distribution \cite{mitchell1968permanence,dufresne2004log}, in agreement with our observation.

In the SLM, the expression for $\beta_i$ is given by Eq.~\eqref{eq:pert-beta-0}, so some variability can be attributed to the noise and growth times \cite{Zaoli2022_BetaDiversity}. This variability is constrained though, because $\tau_iw_{ii}<2$ in order to ensure $\beta_i>0$ (this effect can be also observed in Fig.~1\textbf{b} of Ref.~\cite{Zaoli2022_BetaDiversity}, where $\tau_iw_{ii}$ takes values in the same narrow range). In contrast, the SLVM is free from this limitation because it can reproduce the observed variation in the shape factors even if the product $\tau_iw_{ii}$ is the same for all species. This versatility is a consequence of species interactions [see Eq.~\eqref{eq:pert-beta-1}].


So, can Taylor's law still be considered a valid macroecological pattern for microbiomes? Fig.~\ref{fig:5}\textbf{a} suggests an affirmative answer, but only if it is understood as a pattern `on average', rather than as a categorical law. As a matter of fact, the perturbative expansion of $\beta_i$, Eqs.~\eqref{eq:pert-beta-0} and \eqref{eq:pert-beta-1}, is particularly enlightening in this respect. Assuming that the product $w_{ii}\tau_i$ is the same for all species, Taylor's law becomes exact in the limit of very weak interactions. However, deviations are to be expected when interactions are strong, Eq.~\eqref{eq:pert-beta-1}. In this light, Taylor's law becomes the lowest-order approximation of a more general pattern.


\section{Taylor's law as a Fluctuation-Growth relation}\label{sec:speculation}

The formulation of our model in terms of Langevin equations makes it tempting to draw analogies with a paradigmatic physical system often described with a similar formalism: a Brownian particle moving in one spatial dimension \cite{gardiner2004}.
This is an admittedly ideal system. A further step in complexity would be a system of Brownian particles; to first order, their properties can be well described by the ideal non-interacting case, but at closer look we would find the footprint of interactions: correlations. To keep our discussion intuitive and simple, we will ignore these higher-order effects in both Brownian and microbial systems.

The velocity $v$ of a non-interacting Brownian particle of mass $m$ can be modeled with the Langevin equation
\begin{equation}
m \dot v=-\zeta v + \sqrt{w}\xi(t),
\label{eq:brown}
\end{equation}
where $\zeta$ is the frictional coefficient---as given by Stokes's law $\zeta=6\pi\eta r$, with $\eta$ the viscosity of the fluid and $r$ the radius of the Brownian particle---and $\xi(t)$ is a zero-mean, Gaussian, white noise with correlations $\langle\xi(t)\xi(t^\prime)\rangle= w\delta(t-t^\prime)$.
Writing the corresponding Fokker-Planck equation and calculating the steady state solution, one can derive the following relation for the standard deviation $\sigma$ of the velocity:
\begin{equation}
w={2m \zeta}\sigma^2.
\label{eq:sigma_brown}
\end{equation}
Without any external information provided in addition to the model defined in Eq.~\eqref{eq:brown}, this is all we can say about the fluctuations of the Brownian particle.

The equivalent relations for microbial systems, in the limit of weak interactions, are given by the zeroth order perturbative result in Eq.~\eqref{eq:pert-beta-0}, or, including the first order effect of interactions, by the mean field relation with effective growth rate in Eq.~\eqref{eq:MF-beta-final}. Simplifying notation for the sake of comparison and analogy, we can write (for both zeroth order or mean field, $\tau_i$ is to be understood as it corresponds)
\begin{equation}
w_{ii}=2\tau_i^{-1}\frac{\beta_i^{-1}}{1+\beta_i^{-1}},
\label{eq:wbetai}
\end{equation}
where, using the definition of $\beta_i=(\bar{x}_i/\sigma_i)^2$, we have
\begin{equation}
w_{ii}=2\tau_i^{-1}\frac{{\sigma}_i^2}{\bar{x}_i^2+\sigma_i^2}.
\end{equation}
Again, without any further information beyond the model definition in Eq.~\eqref{eq:mainEquation}, this is all we can say about fluctuations in the microbial population.

However, both for Brownian particles or microbial systems, we have extra information not encoded in the definitions of the corresponding models. For the Brownian particle, the equipartition theorem establishes that $\sigma^2=\langle v^2\rangle=k_B T/m$, where $k_B$ is Boltzmann's constant and $T$ is the temperature of the heat bath in which the system is immersed.
Using this information, we can rewrite Eq.~\eqref{eq:sigma_brown} as
\begin{equation}
w={2 \zeta}k_B T.
\label{eq:sigma_brown_T}
\end{equation}
This is a form of the celebrated Fluctuation-Dissipation relation for the Brownian particle. It links fluctuations of the particle's velocity, determined by $w$, to the energy provided by the environment, in this case the heat bath, characterized by $k_BT$; the proportionality factor depends on the frictional coefficient $\zeta$, that includes the particle's radius and the fluid's viscosity. Note that the relation is independent of $m$, the mass of the particle.

What about microbial populations? Here we do not have an equivalent to the equipartition theorem, but we do have the observational evidence of Taylor's law. Using, as a first approximation, that Taylor's law implies the shape factors of all species to be the same, $\beta_i=\beta$, Eq.~\eqref{eq:wbetai} becomes
\begin{equation}
w_{ii}=2\tau_i^{-1}\frac{\beta^{-1}}{1+\beta^{-1}}.
\label{eq:wbeta}
\end{equation}
This expression is strongly reminiscent of the Fluctuation-Dissipation relation \eqref{eq:sigma_brown_T}. The role of the frictional coefficient $\zeta$, which sets the timescale pushing the particle velocity back to zero (its average), is now played by the growth rate $\tau_i^{-1}$, that sets the timescale driving the abundance back to its average $\bar{x}_i$.
Similarly, the role of temperature is now played by the inverse shape factor $\beta^{-1}$. The main difference is that, while for the Brownian particle, fluctuations increase without bound with $k_B T$, for a microbial species the relation between fluctuation and $\beta^{-1}$ saturates and there is an upper bound given by $w_{ii}=2\tau_i^{-1}$, consistent with the limit that would drive the mean population $\bar{x}_i$ to zero at zeroth order approximation, Eq.~\eqref{eq:pert-mean-0}, or in mean-field, Eq.~\eqref{eq:mean-beta-MF-as-SLM}:strong fluctuations drive species to extinction.

Comparing the fluctuation relations \eqref{eq:sigma_brown_T} and \eqref{eq:wbeta}, it is very tempting to speculate with further parallels.
For the Brownian particle, $k_B T$ represents the environmental input, the energy provided by the heat bath. Analogously, one would speculate $\beta^{-1}$ to be a measure of the environmental input to the microbial population, in this case a complex combination of nutrients, metabolites, and energy sources. We cannot say much of its nature, but the fact of being roughly constant regardless of microbial species hints at some sort of \emph{equipartition} of environmental inputs between different species, just in the same way that energy is equipartitioned between different degrees of freedom of a physical system and a single temperature value characterizes fluctuations for all of them.

The Fluctuation-Dissipation relation \eqref{eq:sigma_brown_T} is independent of the particle mass; likewise, the Fluctuation-Growth relation \eqref{eq:wbeta} is independent of the carrying capacity $K_i$. Both magnitudes play a similar role: while the inverse mass $m^{-1}$ determines how easy it is to accelerate the particle, the carrying capacity $K_i$ determines to which extent a species population can grow.
This analogy can be further stressed by looking at how variances depend on the temperature and the shape factor, respectively. For the Brownian particle
\begin{equation}
\sigma^2 = m^{-1} k_B T,
\label{eq:sigma_m}
\end{equation}
while for a microbial species
\begin{equation}
\sigma_i^2 = K_i^2 \frac{\beta^{-1}}{(1+\beta^{-1})^2}.
\label{eq:sigmai_beta_K}
\end{equation}
We see again that the inverse particle mass $m^{-1}$ and the carrying capacity $K_i$ (or rather its square) both set the scale of the correspondence between heat bath energy $k_B T$ or the Brownian particle, and $\beta^{-1}$ in microbial populations, respectively. Moreover, while for the Brownian particle the variance of the velocity is unbounded, for microbial population it shows a non-monotonic dependence on $\beta^{-1}$, being zero for $\beta^{-1}=0$, peaking for $\beta^{-1}=1$, and returning to zero again as $\beta^{-1}\to\infty$.
The Fluctuation-Growth relation \eqref{eq:wbeta} shows that $\beta^{-1}=0$ corresponds to absence of noise in the system, therefore a null variance was to be expected. However, the same relation gives an increasing relation of the fluctuation coefficient $w_{ii}$ with $\beta^{-1}$. Why, then, a reduction in variance for high values of $\beta^{-1}$, implying high noise? Strong fluctuations, given by high values of $w_{ii}$, push a species closer to extinction, as seen in Eq.~\eqref{eq:pert-mean-0} or Eq.~\eqref{eq:mean-beta-MF-as-SLM}, and extinction implies a vanishing variance. Writing these equations in the simplified notation we are using in this section, i.e.
\begin{equation}
\bar{x}_i =\left(1-\frac{\tau_i w_{ii}}{2}\right) K_i,
\label{eq:mean-K-tau_simple}
\end{equation}
we see that the value $\beta^{-1}=1$ maximizing the variance, corresponds, using Eq.~\eqref{eq:wbeta}, to $w_{ii}=\tau_i^{-1}$, which in turn implies, Eq.~\eqref{eq:mean-K-tau_simple}, $\bar{x}_i = K_i/2$. Intriguingly, Ref.~\cite{grilli2020macroecological} finds that, for very different ecosystems, the values of $\beta$ are of order 1, ranging from 0.2 in seawater to 3.2 in feces. If $\beta^{-1}$ somehow corresponds, as speculated, with environmental input to the microbial system, akin to energy input in physical systems, does its optimization towards $\beta^{-1}=1$ in ecological systems hint at adaptation to utilize resources? Values of $\beta^{-1} \ll 1$ would imply low effect of environmental input to the system, Eq.~\eqref{eq:wbeta}, while $\beta^{-1} \gg 1$ imply strong fluctuations that increase the possibility of species extinction, therefore reducing abundance variance as shown in Eq.~\eqref{eq:sigmai_beta_K}.

Summarizing, mathematical analogies suggest that the observation of Taylor's law in microbial ecological systems may be evidence of an {\it equipartition} relation for environmental resources between different species. This connection remains pure speculation, and further work, out of the scope of this paper, would be necessary to set a basis for such a relation.

\section{Discussion}\label{sec:discussion}

Over the past decades, the breakthrough of metagenomics has sparked an active pursuit to develop models of microbial communities that match the newfound empirical richness. In this endeavor, significant steps have been taken, and different models of population dynamics capable of replicating macroecological patterns of diversity and abundance have been developed \cite{grilli2020macroecological, Ho2022, Descheemaeker2020, Descheemaeker2021, sireci2023environmental, Camacho2024}. These models are typically explored through numerical simulations. Analytical treatments are either scarce or lacking. The tension persists between the growing wealth of empirical data and the analytical challenges in translating these insights into comprehensive, analytically tractable models for understanding the intricacies of microbial communities.

Addressing this challenge, our study introduces an analytical framework designed to explore species abundances and their distribution probabilities, which have recently been experimentally measured and exhibit the shape of gamma distributions \cite{grilli2020macroecological}. The fundamental idea underlying our analysis is to simultaneously address both, biotic forces---primarily represented by species interactions---and abiotic forces---that incorporate, e.g., the effects of the surrounding environment. This marks a reversal of the trend compared to efforts in recent years, which focused on disentangling the various components of dynamics rather than studying their interplay. 


Far from being a mere mathematical exercise, the joint analysis of biotic and abiotic forces is relevant because their combined effects determine whether the system reaches an equilibrium or not. Specifically, one of the key conclusions of our work is that microbial communities generally operate out of equilibrium; communities at equilibrium are unable to generate the correlations between abundances observed in genomic data.

This result provides valuable insights well beyond our model, elucidating, for example, certain aspects of consumer-resource models. Indeed, it has been recently shown that these models predict regimes in which correlations between abundances are absent \cite{sireci2023environmental}. Our analysis explains this fact by revealing that these models are in equilibrium (Sec.~\ref{sec:CR}).

Our work goes past this specific instance and delves into the broader study of non-equilibrium dynamics. As solving the Fokker-Planck equation exceeds our current mathematical skills, we focus on the simpler problem of finding the probability distribution for the abundance of a single species, and propose an approximation leading to an analytical solution for this distribution. What we learn from it is that the empirically observed gamma distribution serves as an effective approximation, performing exceptionally well. Importantly, the parameters of the gamma distribution, such as the shape factors $\beta$, significantly differ from the predictions of the non-interacting model (the SLM) and exhibit a non-trivial variability due to interactions.

The discussion on shape factors stands out as a primary contribution of our work. Firstly, it is essential to highlight that the empirical analysis of nine real biomes reveals that these variables cover a wide range of values, and their distribution significantly deviates from peaking at a constant value (Fig.~\ref{fig:6}), as previously discussed in the non-interacting framework \cite{grilli2020macroecological}. To better understand this behavior, we introduce a perturbative expansion. Perturbative expansions serve as a robust analytical tool, effectively breaking down complex ecological systems into simpler, solvable components (Sec.~\ref{sec:perturbative}). This method facilitates the precise assessment of how different ecological forces contribute to the overall dynamics. Our key conclusion from this analysis is that interspecies interactions alone allow us to reproduce the non-trivial distribution of $\beta$s that is also observed in empirical data. It is true that this variability could also be explained by fine-tuning the values of $\tau_i w_{ii}$---something that has been explored e.g.~in Ref.~\cite{Zaoli2022_BetaDiversity}. This contrived explanation is unsatisfactory though, because it explains really nothing---we must justify why the growth factors have this particular distribution to have a good argument. What is interesting of our finding is that we can explain the variability in the shape factors without any fine-tuning of the interactions. This leads us to think that interactions are the truly relevant ingredient behind this variability. In this regard, our result further substantiates the significant role played by interactions in microbial dynamics \cite{Camacho2024}.

Above all, our analysis of shape factors sheds light on the true nature of Taylor's law (Sec.~\ref{sec:taylor}). A crucial prediction of our work is that the linear relationship commonly presented in recent literature is merely a lowest-order approximation, affected by substantial fluctuations. Furthermore, we find that these fluctuations adhere to a universal distribution across different biomes, revealing a previously unidentified macroecological pattern.

Nevertheless, it is important to remark that Taylor's law---and its consequence, the equality of shape factors---is not a poor approximation because, despite fluctuations, the distribution of beta values remains relatively peaked. Using Taylor's law as a valid coarse-grained relation, we find that imposing it in microbial systems has a similar effect to imposing energy equipartition in physical systems: it induces a Fluctuation-Growth relation akin to Fluctuation-Dissipation in physics. This observation suggests a plausible physical origin of Taylor's law as an effect of resources allocation between species. Future work should incorporate environmental resources and their fluctuations directly in the models, with the goal of deriving Taylor's law from more fundamental principles.

In conclusion, our work represents a step forward in the mathematical modeling of the microbiome, clarifying and discovering certain aspects of its macroecological laws. Moreover, our findings may have concrete implications extending beyond the microbiome, reaching into the broader domain of population theory. This is supported by the recent study by George et al. \cite{george2023}, which aspires to address the temporal evolution of populations in a unified manner.

\section*{Acknowledgments}

We thank Miguel Angel Mu\~noz for his critical reading of the manuscript and both Miguel Angel Mu\~noz and Matteo Sireci for their discussions and suggestions on the non-equilibrium section. This work has been supported by grants PID2022-141802NB-I00 (BASIC), PID2021-128966NB-I00, and PID2022-142185NB-C21 (PGE), funded by MICIU/AEI/10.13039/501100011033 and by ``ERDF/EU''.

\appendix

\section{Solution of the FPE in equilibrium}
\label{app:FPEequilibrium}

The system of equations~\eqref{eq:gradient-eq} can be solved if, and only if, the right-hand side is a gradient field. This occurs provided
\begin{equation}
\frac{\partial^2\log P}{\partial x_j\partial x_k}=
\frac{\partial^2\log P}{\partial x_k\partial x_j},
\end{equation}
a condition that, applied to \eqref{eq:gradient-eq}, translates into
\begin{equation}
\sum_{k=1}^S \frac{m_{ik}a_{kj}}{\tau_k}=0\ \text{for all $i\ne j$,}
\end{equation}
or in matrix form,
\begin{equation}\label{eq:gradient-matrix}
\mathbf{MT}^{-1}\mathbf{A}=-\mathbf{E},
\end{equation}
where $\mathbf{T}=\diag(\tau_i)$, and $\mathbf{E}=\diag(e_i)$ is an (as yet undetermined) arbitrary diagonal matrix. When \eqref{eq:gradient-matrix} holds, the system of equations \eqref{eq:gradient-eq} becomes
\begin{equation}
\frac{\partial\log P}{\partial x_i}=\frac{\beta_i-1}{x_i}-2e_i.
\end{equation}
The solution of these equations is given by Eqs.~\eqref{eq:factorizedSolution} and \eqref{eq:gammaSolution}, where $\bar{x}_i=\beta_i/2e_i$. 

Functions $p_i(x_i)$ can be normalized provided $\beta_i,\bar{x}_i>0$, which in turn imply $e_i>0$ [justifying the choice of sign in \eqref{eq:gradient-matrix}]. Notice that the shape parameters of the gamma distributions $\beta_i$ are solely determined by $\mathbf{W}$ [c.f.~Eq.~\eqref{eq:betas}]. On the other hand, rewriting Eq.~\eqref{eq:gradient-matrix} as
\begin{equation}\label{eq:interaction-equil}
\mathbf{A}=-\mathbf{TWE}
\end{equation}
we can see that fixing $\mathbf{E}$ determines both the mean abundances $\bar{x}_i$ and the interaction matrix $\mathbf{A}$.

\section{Marginal distribution with interactions}
\label{app:marginal}

There is a more standard way of writing the current \eqref{eq:current}. In components,
\begin{equation}\label{eq:orig-current}
J_i(\vec{x})=F_i(\vec{x})P(\vec{x})
-\frac{1}{2}\sum_{j=1}^S\frac{\partial}{\partial x_j}
\big[D_{ij}(\vec{x})P(\vec{x})\big],
\end{equation}
where the components of the drift $\vec{F}(\vec{x})$ are given by \eqref{eq:mainEquation}.

Let us denote $\vec{v}_i$ the vector $\vec{v}$ where the component $v_i$ has been removed. Then, as 
\[
\nabla\cdot\big[\vec{F}(\vec{x})P(\vec{x})\big]=
\nabla_i\cdot\big[\vec{F}_i(\vec{x})P(\vec{x})\big]
+\frac{\partial}{\partial x_i}\big[F_i(\vec{x})P(\vec{x})\big]
\]
and
\[
\int\limits_{\mathbb{R}_+^{S-1}}d\vec{x}_i\,
\nabla_i\cdot\big[\vec{F}_i(\vec{x})P(\vec{x})\big]=0
\]
because all ``surface'' terms vanish due to the boundary conditions, then
\[
\int\limits_{\mathbb{R}_+^{S-1}}d\vec{x}_i\,
\nabla\cdot\big[\vec{F}(\vec{x})P(\vec{x})\big]
=\frac{\partial}{\partial x_i}
\int\limits_{\mathbb{R}_+^{S-1}}d\vec{x}_i\,
\big[F_i(\vec{x})P(\vec{x})\big].
\]
Substituting $F_i(\vec{x})$ [c.f.~Eq.~\eqref{eq:mainEquation}] into the latter integral and factorising $P(\vec{x})=P_i(\vec{x}_i|x_i)p_i(x_i)$, we end up with
\[
\int\limits_{\mathbb{R}_+^{S-1}}d\vec{x}_i\,
\big[F_i(\vec{x})P(\vec{x})\big]=f_i(x_i)p_i(x_i),
\]
where
\begin{align}
\label{eq:drift-i}
&f_i(x_i)=\frac{x_i}{\tau_i}
\left[1+\sum_{i=1}^Sa_{ij}\bar{x}_j(x_i)\right], \\
\label{eq:cond-av}
&\bar{x}_j(x_i)=\int\limits_{\mathbb{R}_+^{S-1}}d\vec{x}_i\,
x_jP_i(\vec{x}_i|x_i).
\end{align}
Note that $\bar{x}_i(x_i)=x_i$. As for the diffusion term,
\[
\int\limits_{\mathbb{R}_+^{S-1}}d\vec{x}_i\,
\frac{\partial^2}{\partial x_i\partial x_j}
\big[D_{ij}(\vec{x})P(\vec{x})\big]=0, \quad j\ne i,
\]
whereas
\[
\int\limits_{\mathbb{R}_+^{S-1}}d\vec{x}_i\,
\frac{\partial^2}{\partial x_i^2}
\big[D_{ii}(\vec{x})P(\vec{x})\big]=\frac{\partial^2}{\partial x_i^2}
\big[D_{ii}(\vec{x})p_i(x_i)\big].
\]

\section{Numerical details}
\label{app:Random-matrix-generation}

\subsection{Langevin's equations}

Simulations of Langevin's equations~\eqref{eq:mainEquation} were carried out using an Euler-Maruyama integration scheme \cite{Toral2014} with time step $\Delta t$. In all cases, growing times $\tau_i$ were set equal for all species ($\tau_i=\tau$), and environmental noise ($\mathbf{W}$) and interaction ($\mathbf{A}$) matrices were randomly chosen as described next.

\subsection{Noise and interaction matrices}

Noise matrices $\mathbf{W}$ must be symmetric, positive definite. In order to randomly generate one such matrix we factor it as $\mathbf{W}=\mathbf{U}\boldsymbol{\Lambda}\mathbf{U}^T$, where $\mathbf{U}$ is an $S \times S$ orthogonal matrix ($\mathbf{UU}^T=\mathbf{U}^T\mathbf{U}=\mathbf{I}$) and $\boldsymbol{\Lambda}$ is a diagonal matrix whose diagonal elements are random, non-negative real numbers. Matrix $\mathbf{U}$ was generated by randomly sampling from a Haar distribution (using the Python function \texttt{ortho\_group} from the SciPy package \cite{2020SciPy-NMeth}). The diagonal elements of $\Lambda$ have been drawn from a uniform distribution $U[0,1]$ (but different probability distributions lead to similar results).

The diagonal elements of the interaction matrices $\mathbf{A}$ were either set to $-1$ or chosen as $a_{ii}=-1/K_i$, where the carrying capacities $K_i$ were sampled from a log-normal distribution $\logN(\mu_l,\sigma_l)$. Parameters $\mu_l$, $\sigma_l$ were tuned so as to align with empirical data for the distribution of mean abundances \cite{Camacho2024}. As for the off-diagonal coefficients, all were set to zero except for a randomly selected fraction $C$ of them (`connectance'). These nonzero elements were drawn from a normal distribution N$(0,\sigma_n)$. For a matrix $\mathbf{A}$ to be considered biologically meaningful and be used in simulations, it must be feasible and stable. Feasibility requires that the stationary abundances of all species are positive. Stability requires that all the eigenvalues of $\mathbf{A}$ have negative real parts \cite{Allesina2012}.

\subsection{Stationary Fokker-Planck equation}

We solved numerically the stationary FPE for a system of two species ($S=2$) using a centered-differences scheme on a square mesh with $N_{x_1}=N_{x_2}$ nodes and spatial length $L_{x_1} = L_{x_2}$ and null boundary conditions. The stationary solution of the FPE is the eigenfunction of the Fokker-Planck differential operator associated to the eigenvalue $\lambda=0$. However, after discretizing the FPE, $\lambda=0$ is no longer an eigenvalue of the resulting matrix. Hence, we obtained the stationary solution as the eigenvector associated to the eigenvalue with smallest $|\lambda|$. To find that eigenvector, we implemented an inverse-power iterative method \cite[Sec.~4.16.3]{engeln2013numerical} solving the linear system through a variant of Thomas's algorithm \cite[Sec.~7.3.2]{engeln2013numerical} to take advantage of the block-tridiagonal structure of the matrix.

\bibliography{bibliography}

\end{document}